\documentclass[10pt,letterpaper,twocolumn]{article} 

\usepackage{OL}
\usepackage{hyperref}
\usepackage{amsmath}
\usepackage{graphicx}

\usepackage{amssymb}
\usepackage{amsfonts}
\usepackage{array}

\usepackage{graphicx}
\usepackage{subfigure}

\begin{document}

\twocolumn[ 

\title{Twin Families of Bisolitons in Dispersion Managed Systems}
\author{Ildar Gabitov$^{1,3}$, Robert Indik$^{1}$, Pavel Lushnikov$^{2}$, Linn Mollenauer$^{1}$, Maxim Shkarayev$^{1}$}
\address{$^{1}$Department of Mathematics The University of Arizona 617 N.
Santa Rita Ave.  Tucson, AZ 85721\\
$^{2}$Department of Mathematics \& Statistics,  MSC03 2150,  University of New Mexico
      Albuquerque, NM 87131-1141 \\
     $^3$ Landau Institute for Theoretical Physics, Kosygin St. 2,
  Moscow, 119334, Russia }

\maketitle

\begin{abstract}
We calculate bisoliton solutions using a slowly varying stroboscopic
equation. The system is  characterized in terms of a single
dimensionless parameter.  We find two branches of solutions and
describe the structure of the tails for the lower branch solutions.
\end{abstract}
\ocis{060.2310, 190.4370, 060.2330, 060.5530, 190.4380.} ]

Bisolitons in optical fiber lines with dispersion-management, were
first discovered using computer modeling~\cite{Maruta_02} and later
experimentally~\cite{Mitschke_05}.  Bisolitons can be viewed as a
two component soliton molecule. In numerical simulations, they are
stable over long propagation distances and, if perturbed, oscillate
about equilibrium. We investigate the structure of such pulses
assuming that fiber losses are completely compensated and
propagation of pulses through optical fiber in a dispersion managed
system is governed by the nonlinear Schr\"odinger equation:
\begin{equation}
i u_z + d(z) u_{\text{tt}} + \gamma |u|^2 u = 0 \label{NLS}
\end{equation}
where $u=u(t,z)$ is the slowly varying envelope of the
electromagnetic field inside the fiber. We consider a simple case of
a piecewise constant dispersion function $d(z)$, where a fiber span
of length $z_{dm}/2$ with normal dispersion alternates with
equal-length spans of anomalous dispersion fiber.  The function
$d(z)$ can be represented as a sum of an oscillating part
$\tilde{d}(z)$ and a residual dispersion $d_0$ such that $d(z) = d_0
+ \tilde{d}(z)$. Here $\langle \tilde{d}(z) \rangle = 0$;
$\tilde{d}(z)=d_1,$ if $0\le z\le z_{dm}/2$, and $\tilde{d}(z)=-d_1$
if $z_{dm}/2 \le z \le z_{dm}$.  In this system, the characteristic
length of the nonlinearity is $z_{nl}\sim 1/|P|^2$,  where $P$ is
the peak power of the bisoliton, while the characteristic length of
the residual dispersion is $z_{d0}\sim \tau^2/d_0$, where $\tau$ is
the pulse width. If the period of the dispersion map $z_{dm}$ is
much smaller than $z_{nl}$ and $z_{d0}$, then the spectrum $\hat{u}$
of the solution to Eq.~(\ref{NLS}) is a slowly varying function of
$z$ on the scale $z_{dm}\ll z_{nl},z_{d0}$ and can be represented as
\begin{equation}
\hat{u}= q(\omega,z) \exp \left( -i \omega^{2} \int_{z_{dm}/4}^{z}
  \tilde{d}(z')dz' \right) \label{fast_slow}
\end{equation}
The exponential term captures the fast (in $z$)  phase and
$q(\omega,z)$ captures the slow amplitude dynamics of the spectral
components. As has been shown~\cite{GT_96}, the evolution of the
spectral components at leading order can be described by
\begin{eqnarray}
  i q_z(\omega) -  d_0\omega^2 q(\omega) +\gamma
   R(q(\omega),\omega)=0,\quad\mbox{where}\nonumber\\
R(q(\omega),\omega)=\frac{1}{(2\pi)^2} \int \frac{\sin(s \Delta /2)}{s
\Delta /2}q(\omega_1)q(\omega_2)q^*(\omega_3) \nonumber\\
  \times  \delta(\omega_1 + \omega_2
- \omega_3 - \omega) d \omega_1 d \omega_2 d \omega_3 \label{Slow}
\end{eqnarray}
Here $s \equiv z_{\text{dm}} d_1/2$  is dispersion map strength and
$\Delta \equiv \omega_{1}^2+\omega_{2}^2-\omega_{3}^2-\omega^2$.  We
determine a shape of a bisoliton solution following earlier work by
P.L~\cite{Lushnikov01}. If a solitary wave solution with phase
period $\lambda^{-1}$ has the form $q(\omega) = A(\omega)e^{i\lambda
z}$, then amplitude $A(\omega)$ evolves according to the integral
equation:
\begin{equation}
-\lambda A -  d_0\omega^2  A + \gamma R(A(\omega),\omega) = 0
\end{equation}
Rescaling  variables $t = \tau_0\tau$,  $\omega =  \Omega /\tau_0$
and $ A(\omega) = a \varphi(\Omega)$,  where  $\tau_0 = s^{1/2}$,
$a = 2 \pi (s \lambda /\gamma)^{1/2}$, results in a dimensionless
equation
\begin{eqnarray}
   - \left( \ 1+ \bar{d_{0}}\Omega^2\right)\varphi(\Omega)+
\bar{R}\left(\varphi(\Omega),\Omega\right)= 0,\quad\mbox{where} \nonumber \\
\bar{R}(\varphi(\Omega),\Omega) = \frac{1}{(2\pi)^2}\int \frac{\sin(
\bar{\Delta} /2)}{\bar{\Delta}/2} \varphi(\Omega_1)
\varphi(\Omega_2) \varphi^*(\Omega_3) \nonumber \\
\times \delta(\Omega_1 + \Omega_2 - \Omega_3 - \Omega)\text{d$\Omega_1$}
\text{d$\Omega_2$}\text{d$\Omega_3$}, \label{slow_dimensionless}
\end{eqnarray}
which depends on a single parameter $\bar{d_{0}}= d_{0}/(s \lambda)$.
Here $\bar{\Delta} = \Omega_1^2 + \Omega_2^2 -
\Omega_3^2 - \Omega^2$.

We study the structure of bisolitons as a function of $\bar{d_{0}}$.
Following the experimental work~\cite{Mitschke_05} we consider
antisymmetric solutions of Eq.~(\ref{slow_dimensionless}).  To solve
this integral equation we use the iterative procedure:
\begin{equation}
\varphi_{n+1}(\Omega) =\mathbb{P}_{\text{odd}}\left( Q_n^{3/2}
\frac{\bar{R}(\varphi_n(\Omega),\Omega)}{1 +
\bar{d_0}\Omega^2}\right)\label{scheme}
\end{equation}
Here $\mathbb{P}_{\text{odd}}(f(x)) = (f(x)-f(-x))/2$ is a
projection operator onto the set of odd functions and $\hat{F}^{-1}$
is an inverse Fourier transform.  A modified Petviashvili
stabilizing factor~\cite{Petviashvili} $Q_n$
\begin{equation*}
  Q_n \equiv  \left[ {\hat{F}^{-1}\left[ \varphi_n(\Omega) \right]}
\bigg/
{\hat{F}^{-1} \left[ \frac{\bar{R}(\varphi_n(\Omega),\Omega)}{1 + \bar{d_{0}}\Omega^2}
\right]} \right]_{\tau = 0.5}
\end{equation*}
allows the scheme to avoid trivial solutions  $\varphi = 0$.  The
most costly part of this iterative procedure is evaluation of
$\bar{R}$, which involves a triple integral. To expedite evaluation
of these integrals we used the procedure described by
P.L~\cite{Lushnikov01}. It should be noted that the bisoliton
solution of~Eq.(\ref{slow_dimensionless}) represents the unchirped
pulse shape at the middle of each span with positive dispersion.

We begin by studying the solutions for the parameter
$\bar{d_{0}}=0.067$ (a realistic value for communication
systems~\cite{MG2006}), choosing $\varphi_0$ as a sum of two shifted
real valued  Gaussian functions with opposite signs. The iteration
procedure converges to the fixed point as the Petviashvili factor
$Q_n$ approaches 1. The iteration is stopped when $|Q_n -
1|<10^{-5}$. The value of $\bar{d_{0}}$ was then varied in small
increments.  We use the solution found for the nearby $\bar{d_{0}}$
as the initial ``guess'' for the next  value of $\bar{d_{0}}$.
\begin{figure}[!t]
\includegraphics[width=3.4in]{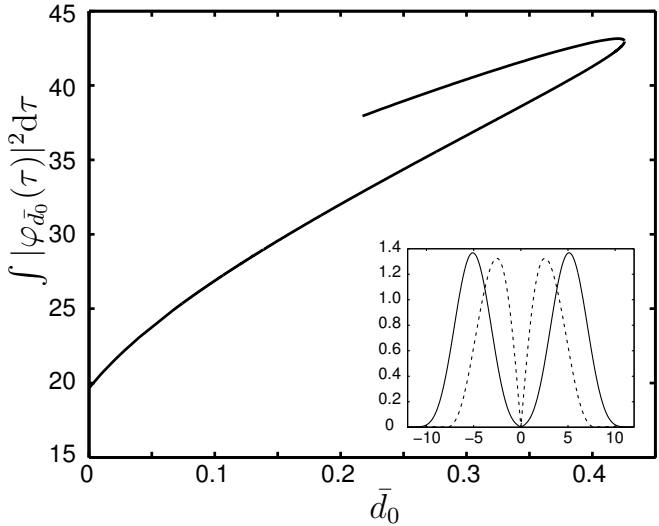}
      \caption{ Bisoliton energy as two valued function of dimensionless residual dispersion
$\bar{d_{0}}$. Solid and dashed lines on insert correspond to the
upper and lower branch bisolitons with $\bar{d_{0}}=0.256$.
Horizontal and vertical axes on insert correspond to dimesionless
time and amplitude.  }
      \label{sen}
\end{figure}
Fig.~\ref{sen} represents  bisoliton energy as function of
$\bar{d_{0}}$. Remarkably, this is a multiple valued function with
two branches. Calculation of solutions on this second branch
required solution of the Arnoldi-Lanczos approximation problem for
the linearization of iterration operator. Its limit point is located
in the vicinity of $\bar{d_{0}}_{bf} \simeq 0.426$. With all other
parameters fixed, smaller values of $\bar{d_{0}}$ correspond to
smaller values of residual dispersion. Therefore, the limit point
corresponds to the largest value of $d_{0}$ for which bisolitons are
supported. According to our calculations, for values of $\bar{d_{0}}
> \bar{d_{0}}_{\text{bf}}$ bisolitons will fail to exist and we will
only observe a pair of interacting dispersion managed solitons that
are not bound, not a bisoliton. The insert to the figure shows that
the higher energy bisoliton is wider, with greater separation and
broader shape.

Direct numerical simulations demonstrate stability of both the lower
and the upper-branch bisoliton solutions over realistic distances
(300 periods). For very long propagation distances, the upper branch
showed signs of instability.
Additionally, the pair of pulses composing the solution tends to
stay bound whenever the pulses are pulled apart. The separation
between the pulses spread apart in this manner oscillates about the
separation for the bisolitonic solution.

\begin{figure}[t!]
\includegraphics{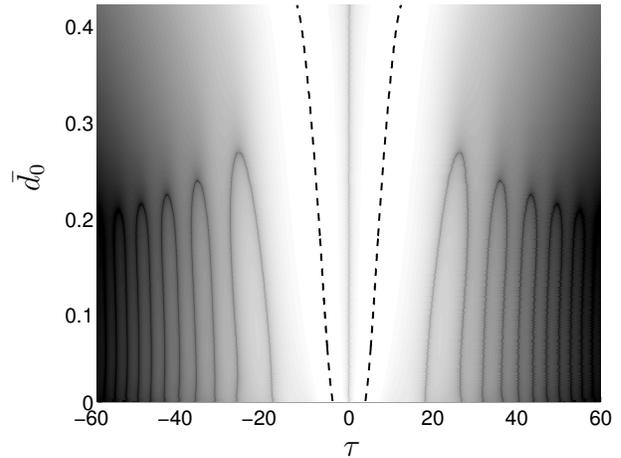}
      \caption{The contour plot of $\log (|\varphi|)$ as a
function of time variable $t$ and a parameter $\bar{d_{0}}$. The
dotted lines represents the peak amplitudes of solutions.}
      \label{slog}
\end{figure}

\begin{figure}[]
\includegraphics[width=3.4in]{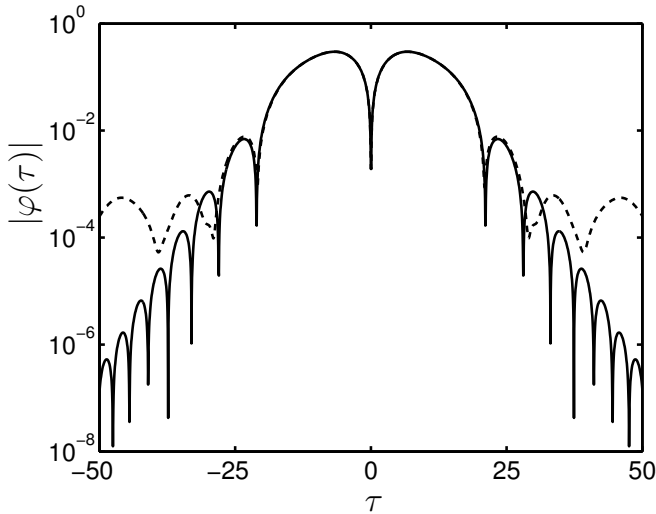}
      \caption{Intensity profiles of a bisoliton with $\bar{d_{0}}=0.167$: solution of Eq.~(\ref{slow_dimensionless}) (solid), and a result of propagation this solution over 300 periods through the line with $s=1.5$, $d_{0} = 0.0125$, and $\lambda=0.05$ (dashed)}
      \label{sbar6}
\end{figure}

In the remainder of this paper, we will discuss the structure of the
lower branch solutions. A later paper will present details about the
upper branch solutions. The logarithmic density profile of lower
branch solutions' amplitude for a range of values of $\bar{d_{0}}$
is shown in Fig.~{\ref{slog}}. There  lighter shades of gray
correspond to a higher value of the amplitude. The black lines
correspond to zero values. The dashed lines indicate where the
solutions have their maxima. A horizontal slice of this plot gives
an amplitude profile for a fixed value of $\bar{d_{0}}$. For
example, for a value of $\bar{d_{0}}=0.167$ the logarithm of
amplitude is represented on Fig.~{\ref{sbar6}}. The dashed line on
the same graph represents the result of direct numerical simulations
of Eq. (\ref{NLS}) after 300 dispersion map periods. A solution of
equation~\ref{slow_dimensionless} provided one boundary condition
for this simulation (the launched pulse).

As we see in Fig.~{\ref{slog}} bisoliton tails change sign for
values of $\bar{d_{0}} < \bar{d_{0}}_{cr}$ ( $\bar{d_{0}}_{cr}\simeq
0.269$ ). As the value of the dimensionless residual dispersion
becomes greater than the critical value $\bar{d_{0}}
> \bar{d_{0}}_{cr}$, the phase of the tails remains unchanged, and the amplitude
approaches an exponentially decaying function.
Our equation Eq.~(\ref{slow_dimensionless}) reduces the many
physical parameters from Eq.~(\ref{NLS})to a single parameter.  We
have computed the bisolitons for a range of parameter values, and
those computed solutions can be used to write solutions of the
original physical system
\begin{equation}
u(t,z_{dm}/4+mz_{dm})=  \sqrt{\frac{\lambda }{\gamma}}
\hat{F}^{-1} \left.\left[ \varphi_{\bar{d_{0}} = \frac{d_0}{\lambda
s}}\left(\Omega\right) \right ] \right|_{\tau = t/\sqrt{s}}
\label{solution}
\end{equation}
where it was convenient to specify the solution at
$z_{dm}/4+mz_{dm}$ because it is chirp free at this point.

\begin{figure}[htb!]
\subfigure[]{\label{s6}\includegraphics[width=3.4in]{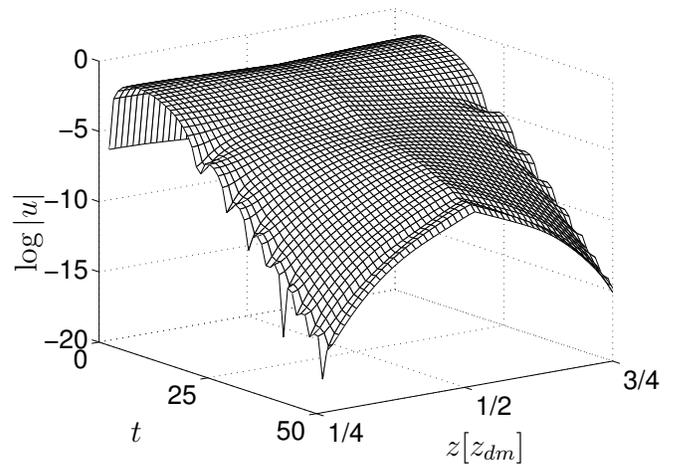}}
\subfigure[]{\label{s24}\includegraphics[width=3.4in]{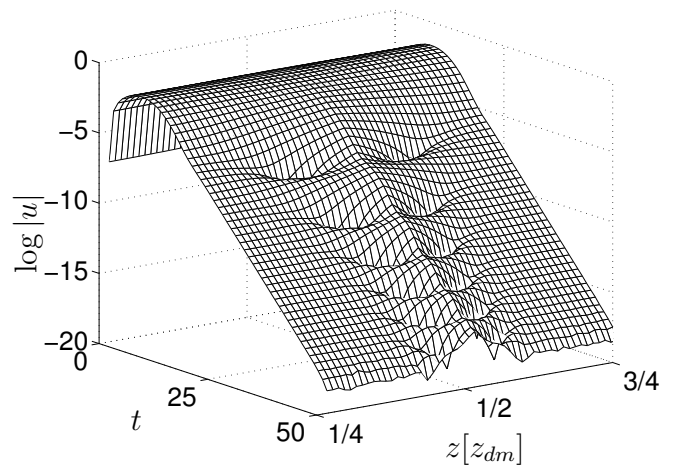}}
      \caption{Waveforms of a logarithm of the amplitude for bisoliton propagation 
through DM system with $s=1.5$, $d_0 = 0.0125$, $\gamma=1$ (a) $\lambda = 0.05$ and (b) $\lambda = 0.02$. The soliton magnitude $|u|$ is symmetric. 
Bisolitons are propagated over half of the period, from $z_{dm}/4$ to $3z_{dm}/4$}
      \label{NLS_propogation}
\end{figure}

We use $u(t,z_{dm}/4)$ of Eq.~(\ref{solution}) as the initial
condition in direct numerical simulation of Eq.~(\ref{NLS}) to study
the dynamics of lower-branch bisoliton solutions for different
values of $\bar{d_{0}}$. In particular, we compare temporal-spatial
behavior of solutions corresponding to  values of $\bar{d_{0}}_{cr}
< \bar{d_{0}} <\bar{d_{0}}_{bf} $ and $ 0 < \bar{d_{0}} <
\bar{d_{0}}_{cr}$ over a map period. We consider a system with
$s=1.5$ and mean dispersion $d_0 = 0.0125$. Fig.~\ref{s6} and
Fig.~\ref{s24} represent propagation of the initial pulse with
$\bar{d_{0}}=0.417$ and $\bar{d_{0}}=0.167$ respectively. For such
choices of the system parameters, the values of phase periods
$\lambda$ must be 0.05 and 0.02.

The bisolitons for values of $\bar{d_{0}}$ above the critical value
have no zeros other than at $\tau=0$ in their unchirped state, while
if $\bar{d_{0}} < \bar{d_{0}}_{cr}$ the solution will have an
increasing number of zeros with smaller $\bar{d_{0}}$. The example
shown in Fig.~\ref{s24} shows that for larger values of $d_0$ the
local minima which in part (a) are at $z=0$   split into pairs of
minima, which are  "shifted"  towards $z_{dm}/2$. In fact, comparing
the dynamics of $\bar{d_{0}}=0.417$ and $\bar{d_{0}}=0.167$
bisolitons indicates that the larger the value of $\bar{d_{0}}$ the
more the minima will shift from the narrowest states (the valleys)
of the pulse to its broadest state (the ridges).

In conclusion, we have used a slowly varying stroboscopic equation
to calculate bisoliton solutions with well resolved tails.  This
equation can be rescaled so that it has a single dimensionless
parameter $\bar{d}_0$.  We have found a range of $\bar{d}_0$ such
that there are two bisoliton solutions for each value of
$\bar{d}_0$.  In addition, the structure of the tails for the lower
branch solutions was described in terms of the value of $\bar{d}_0$.

We would like to acknowledge many helpful discussions and useful
suggestions contributed by M. Stepanov. This work was supported in
part by Los Alamos National Laboratory under an LDRD grant, the
National Nuclear Security Administration of the U.S. Department of
Energy under Contract \# DE-AC52- 06NA25396, by the DOE Office of
Science Advanced Scientific Computing Research (ASCR) Program in
Applied Mathematics Research, as well as Proposition 301 funds from
the State of Arizona.

\end{document}